\documentclass[%
 aip,
 amsmath,amssymb,
 reprint,%
floatfix,
]{revtex4-1}

\usepackage{graphicx}% Include figure files
\usepackage{dcolumn}% Align table columns on decimal point
\usepackage{bm}% bold math
\usepackage[dvipsnames]{xcolor}
\usepackage[mathlines]{lineno}% Enable numbering of text and display math
\usepackage[utf8]{inputenc}
\usepackage[T1]{fontenc}
\usepackage{mathptmx}
\usepackage[english]{babel}
\usepackage{wasysym}
\usepackage{fdsymbol}

\makeatletter
\def\bbl@set@language#1{%
  \edef\languagename{%
    \ifnum\escapechar=\expandafter`\string#1\@empty
    \else\string#1\@empty\fi}%
  \@ifundefined{babel@language@alias@\languagename}{}{%
    \edef\languagename{\@nameuse{babel@language@alias@\languagename}}%
  }%
  \select@language{\languagename}%
  \expandafter\ifx\csname date\languagename\endcsname\relax\else
    \if@filesw
      \protected@write\@auxout{}{\string\select@language{\languagename}}%
      \bbl@for\bbl@tempa\BabelContentsFiles{%
        \addtocontents{\bbl@tempa}{\xstring\select@language{\languagename}}}%
      \bbl@usehooks{write}{}%
    \fi
  \fi}
\newcommand{\DeclareLanguageAlias}[2]{%
  \global\@namedef{babel@language@alias@#1}{#2}%
}
\makeatother

\DeclareLanguageAlias{en}{english}

\begin{document}

\preprint{AIP/123-QED}

% Guidelines from APL Materials > no real length restrictions, just recommended length
% \onecolumngrid
% \textcolor{red}{APL -- 3000 words, including figure and table captions. \textbf{As of 11/27/2020: 2773 words}}

\title{Tailoring Superconducting Phases Observed in Hyperdoped Si:Ga for Cryogenic Circuit Applications}
% Force line breaks with \\

\author{K. Sardashti}
\affiliation{%
Department of Physics, New York University, New York, NY 10003%\\This line break forced% with \\
}%

\author{T. Nguyen}
\affiliation{%
Department of Physics, New York University, New York, NY 10003%\\This line break forced% with \\
}%
\affiliation{ 
Department of Physics, City College of New York, City University of New York, New York, NY 10031%\\This line break forced with \textbackslash\textbackslash
}%

\author{M. Hatefipour}
\affiliation{%
Department of Physics, New York University, New York, NY 10003%\\This line break forced% with \\
}%

\author{W. L. Sarney}
\affiliation{%
CCDC U.S. Army Research Laboratory, Adelphi, MD 20783 USA%\\This line break forced% with \\
}%

\author{J. Yuan}
\affiliation{%
Department of Physics, New York University, New York, NY 10003%\\This line break forced% with \\
}%

\author{W. Mayer}
\affiliation{%
Department of Physics, New York University, New York, NY 10003%\\This line break forced% with \\
}%

\author{K. Kisslinger}
\affiliation{%
Center for Functional Nanomaterials, Brookhaven National Laboratory, Upton, NY 11973%\\This line break forced% with \\
}%

\author{J. Shabani}
\affiliation{%
Department of Physics, New York University, New York, NY 10003%\\This line break forced% with \\
}%

\date{\today}% It is always \today, today,

\begin{abstract}
Hyperdoping with gallium (Ga) has been established as a route to observe superconductivity in silicon (Si). The relatively large critical temperatures (T$_{\rm c}$) and magnetic fields (B$_{\rm c}$) make this phase attractive for cryogenic circuit applications, particularly for scalable hybrid superconductor--semiconductor platforms. However, the robustness of Si:Ga superconductivity at millikelvin temperatures is yet to be evaluated. Here, we report the presence of a reentrant resistive transition below T$_{\rm c}$ for Si:Ga whose strength strongly depends on the distribution of the Ga clusters that precipitate in the implanted Si after annealing. By monitoring the reentrant resistance over a wide parameter space of implantation energies and fluences, we determine conditions that significantly improve the coherent coupling of Ga clusters, therefore, eliminating the reentrant transition even at temperatures as low as 20~mK.
\end{abstract}

\maketitle

%################################
% INTRODUCTION
%################################

Superconducting silicon (Si) is considered a key to realization of all-Si hybrid superconductor--semiconductor (S-Sm) devices that are ideal for highly scalable quantum circuits. \cite{shim_bottom-up_2014, shim_superconducting-semiconductor_2015} Superconductivity in Si has been demonstrated by hyperdoping beyond metal--insulator transition limits via boron (Si:B) \cite{bustarret_superconductivity_2006, marcenat_low-temperature_2010, grockowiak_thickness_2013} and gallium (Si:Ga). \cite{skrotzki_-chip_2010, fiedler_superconducting_2011, heera_silicon_2013} Moreover, hyperdoped Si:B has been successfully integrated into  gate-tunable all-Si Josephson junctions (JJs) \cite{chiodi_proximity-induced_2017} and superconducting quantum interference devices (SQUIDs).\cite{duvauchelle_silicon_2015} Nevertheless, relatively low critical temperature (T$_{\rm c} <$ 800~mK) and magnetic fields (B$_{\rm c} <$ 1~T) for superconducting phases of Si:B would possibly limit their application in quantum circuits, due to susceptibility to thermal and flux noise.

The superconducting phase in hyperdoped Si:Ga may be a plausible alternatives because of its larger T$_{\rm c} \sim$ 6~K and  B$_{\rm c} \sim$ 10~T.\cite{skrotzki_-chip_2010, fiedler_superconducting_2011} But Si:Ga transport properties at temperatures far below its T$_{\rm c}$ have not been explored.  Such temperatures are commonly used for quantum devices in order to minimize the influence of thermal energy ($\sim$ k$_{\rm B}$T) on the quantum two-level systems. Given that the superconductivity in Si:Ga has been attributed to the presence of a few $nm$-thick $Ga$ layer segregated near the top surface, it is possible that a re-entrant resistive phase exists below T$_{\rm c}$ due to disorder and percolation induced by Si or SiO$_{\rm x}$ mixing \cite{fiedler_superconducting_2011, heera_silicon_2013,thorgrimsson_effect_2020, fischer_optical_2013}. Therefore, it is critical to determine the superconducting characteristics  for hyperdoped Si:Ga well below its critical temperature all the way down to mK range.

In this work, we study the influence of processing condition on the superconducting properties of hyperdoped Si:Ga prepared by $Ga^{\rm +}$ implantation. We first demonstrate that the implantation energy of 80~keV, which is  used in all previous studies of Si:Ga superconductivity \cite{skrotzki_-chip_2010, heera_silicon_2013, thorgrimsson_effect_2020}, leads to emergence of a re-entrant insulating state below T$_{\rm c}$. We then tailor the $Ga$ distribution within the implanted region by tuning processing parameters such as implantation energy (E$_{\rm IMP}$), dopant activation temperature (T$_{\rm DA}$) and Ga$_{\rm +}$ fluence ($\rm {\Phi}_{\rm Ga}$). Although superconductivity is observed over a wide parameter space, resistive transition temperature (T$_{\rm c}$) remains within a narrow range of 5.2 -- 7.1~K. This is in agreement with superconductivity stemming from confined crystalline (e.g., $\beta$-phase) or amorphous $Ga$ with varying degrees of coherent coupling \cite{jaeger_threshold_1986, hagel_superconductivity_2002}. Furthermore, we demonstrate conditions where the superconducting network in Si:Ga reaches the connectivity level necessary to maintain dissipationless conductivity at temperatures as low as 20~mK.

In order to prepare hyperdoped Si:Ga we started with undoped Si(100) wafers ($\rho >$~1000~$\Omega.cm$). A 30 $nm$ thick SiO$_{\rm 2}$ cap layer was deposited on each wafer using plasma-enhanced chemical vapor deposition (PECVD). Samples underwent $Ga^{\rm +}$ implantation at ambient temperature with energies E$_{\rm IMP}$ = 25~keV--80~keV, and fluences $\rm {\Phi}_{\rm Ga}$ = 4 -- 6x10$^{\rm 16} cm^{\rm -2}$. Implanted wafers then underwent dopant activation annealing at temperatures (T$_{\rm DA}$) of 300 -- 800~$^{\circ}$C for 1~min in N$_{\rm 2}$ using a rapid thermal annealer.

\begin{figure}[ht!]
\includegraphics[width=0.48\textwidth]{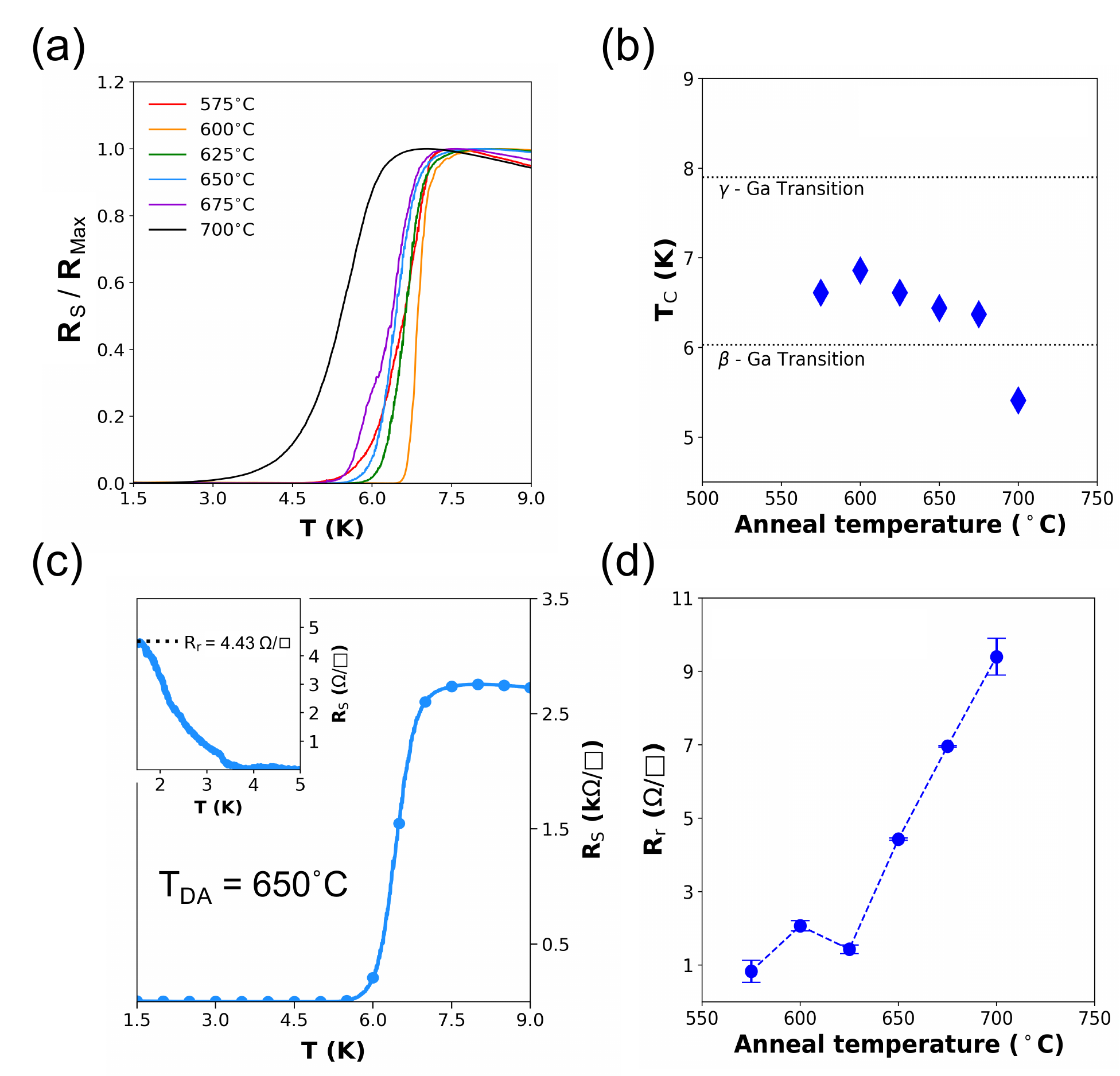}
\caption{\label{fig.1} (a) Sheet resistance R$_{\rm S}$ (normalized to its peak value below 10K, R$_{\rm Max}$) vs temperature for Si:Ga samples with implantation energy E$_{\rm IMP}$ of 80keV. (b) Resistive transition temperatures T$_{\rm c}$ vs anneal temperature T$_{\rm DA}$ for the samples shown in (a). (c) Superconducting transition of a Si:Ga sample with E$_{\rm IMP}$ = 80keV that is annealed at 650$^{\circ}$C. The inset shows transport behavior below 5~K (sub-- T$_{\rm c}$) marking the reentrant resistance (R$_{\rm r}$) as the sheet resistance value at T = 1.5 K. (d) Measured reentrant resistance R$_{\rm r}$  as a function of anneal temperature T$_{\rm DA}$.} 
\end{figure}

\setcitestyle{numbers}
Transport properties were initially evaluated by measuring differential resistance vs temperature down to 1.5~K in Van der Pauw (VdP) geometry (using an Oxford Instruments Teslatron PT measurement system). In Fig. \ref{fig.1}, we show the superconducting characteristics of the Si:Ga chips implanted at E$_{\rm IMP}$ = 80~keV and ${\rm \Phi}_{\rm Ga}$ = 4 $\times$ 10$^{\rm 16}$~cm$^{\rm -2}$; these parameters are adapted from previous reports of Si:Ga superconductivity in Ref. \cite{skrotzki_-chip_2010, fiedler_superconducting_2011, heera_silicon_2013, fischer_optical_2013}.
\setcitestyle{super} Fig.\ref{fig.1}(a) shows the sheet resistance R$_{\rm S}$, normalized to its maximum value below 10~K (R$_{\rm Max}$),  vs temperature for six superconducting samples with T$_{\rm DA}$ increasing from 575 to 700 $^{\circ}$C. The resistive transition temperature T$_{\rm c}$ for the superconducting samples is shown in Fig.\ref{fig.1}b. T$_{\rm c}$ for each R$_{\rm S}$(T) trace is defined as the temperature where  R$_{\rm S}$ = 0.5 R$_{\rm n}$. As dopant activation temperature increases, T$_{\rm c}$ varies $\pm$ 0.8~K around the $\beta$-Ga T$_{\rm c}$ \cite{campanini_raising_2018}. This is consistent with the solid-state precipitation of $Ga$ within the implanted Si:Ga to form a superconducting Ga-rich network.

% %%### Figure 1 ###############################
% %%#######################################################

In order to evaluate the superconducting behavior below T$_{\rm c}$, we examined the temperature dependence of sheet resistance R$_{\rm S}$(T) in 1.5~K -- 6~K temperature range. Fig.\ref{fig.1}c displays the R$_{\rm S}$(T) trace for the sample with T$_{\rm DA}$ = 650~$^{\circ}$C. In the inset we show the emergence of a reentrant resistance as temperature approaches 1.5~K. In order to compare the extent of reentrant behavior between samples, we identified R$_{\rm S}$ at 1.5~K as the characteristic reentrant resistance (R$_{\rm r}$). In Fig.\ref{fig.1}d we show the dependence of R$_{\rm r}$ on T$_{\rm DA}$ for samples E$_{\rm IMP}$ = 80~keV, where the lower T$_{\rm DA}$ typically leads to smaller re-entrant resistance. The dramatic growth in R$_{\rm r}$ above 625~$^{\circ}$C may be attributed to the loss of Ga, leaving poorly-connected $Ga$ networks behind.

% %%### FIGURE 2 ###############################
\begin{figure}[ht!]
\includegraphics[width=0.45\textwidth]{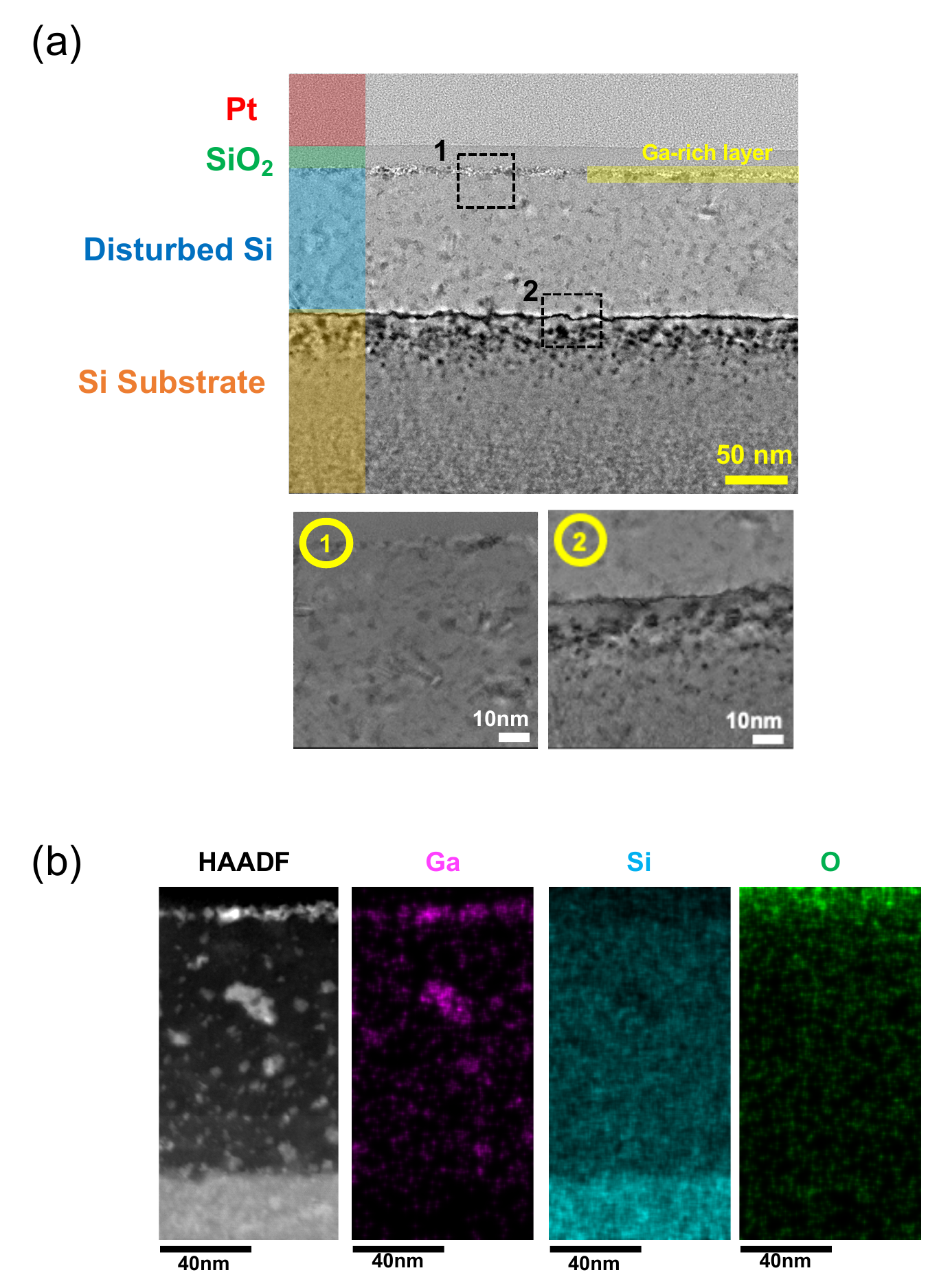}
\caption{\label{fig.2} (a) Transmission electron microscopy (TEM) images of cross-sections prepared on Si:Ga sample with E$_{\rm IMP}$ = 80keV and $\Phi_{\rm Ga}$ = 4$\times 10^{\rm 16}$ cm$^{\rm -2}$, annealed at 700~$^{\circ}$C. The dotted boxes shown in top image are visual guides to outline regions where high-resolution images (1) and (2) are taken. (b) Dark field image and elemental maps for Ga, Si, and O  obtained by EDS on cross-section of the sample shown in (a).}
\end{figure}
% %%#########################################

For the sample with the largest reentrant resistance (i.e., E$_{\rm IMP}$ = 80~keV and T$_{\rm DA}$ = 700 $^{\circ}$C), we investigated the structural characteristics using cross-sectional Transmission Electron Microscopy (TEM). Fig. \ref{fig.2}(a) displays TEM images of the sample cross-section. Up to 100 $nm$ below the SiO$_{\rm 2}$ cap, we observe nano-crystalline Si whose structure was recovered during the post-implantation annealing. Within this region, as well as at the Si/SiO$_{\rm 2}$ interface, a secondary phase is present in the form of $\sim$ 5--10 $nm$ wide clusters. These clusters appear to vary from crystalline to amorphous depending on the location within the implanted region. High-angle annular dark-field (HAADF) images and energy dispersive X-ray spectroscopy (EDS) elemental maps for the same cross-section, shown in Fig.\ref{fig.2}(b), confirm that the clusters are highly Ga-rich. $Ga$ precipitation is highly likely in the Si:Ga system as $Ga$ is only soluble in Si up to 0.1 at.\% at 1000~$^{\circ}$C \cite{olesinski_gasi_1985}. Particularly at elevated temperatures (e.g., 700~$^{\circ}$C) insoluble $Ga$ is expected to precipitate into clusters within the bulk or below the SiO$_{\rm 2}$ barrier. The more closely-packed clusters at Si/SiO$_{\rm 2}$ may form a pseudo-2D $Ga$ thin film to host the superconductivity. However, the layer appears to be granular with regions bridges with few $nm$ wide Si weak-links. Although those weak-links are highly p-doped, their carriers could eventually freeze in large fractions at T $<$ 1~K. This will limit their ability to carry supercurrents through the proximity effect, as their superconductivity eventually vanishes at very low carrier concentration \cite{nishino_carrier-concentration_1986}. Therefore, the coherent coupling coupling between superconducting Ga puddles can eventually be destroyed at near-zero temperatures \cite{kim_observation_2001, ansermet_reentrant_2016}.

\begin{figure}
\includegraphics[width=0.48\textwidth]{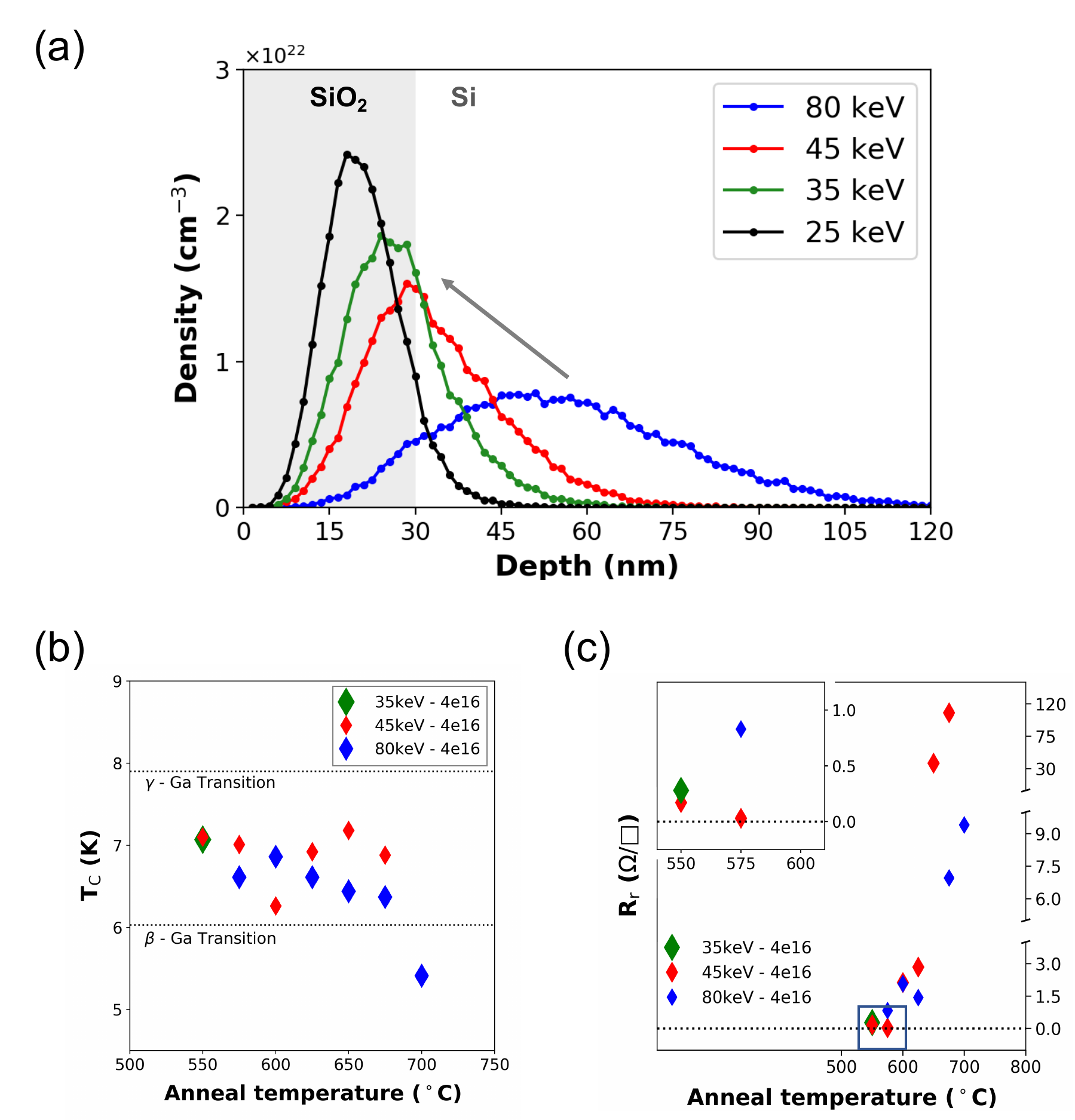}
\caption{\label{fig.3} (a) TRIM simulations of $Ga$ concentration vs depth at various implantation energies E$_{\rm IMP}$ from 25~keV to 80~keV with a fixed fluence ${\rm \Phi}_{\rm Ga}$ of 4 $\times$ 10$^{\rm 16}$\ $\rm cm^{-2}$. In all simulations, a 30~$nm$ thick SiO$_{\rm 2}$ is accounted for. (b) Transition temperatures T$_{\rm c}$ vs anneal temperature T$_{\rm {DA}}$ for all the superconducting Si:Ga samples with E$_{\rm IMP}$ between 35~keV and 80~keV. (c) Reentrant resistance at 1.5~K R$_{\rm r}$ as a function of T$_{\rm DA}$ and E$_{\rm IMP}$. The inset magnifies the region outlined by the box near R$_{\rm r} \sim$ 0 ${\rm \Omega} / \square$. 
}
\end{figure}

To boost connectivity of $Ga$ clusters, we lowered the $Ga^{\rm+}$ implantation energy to less than 80~keV. In Fig.\ref{fig.3}(a), we show the $Ga$ concentration vs depth calculated by the Transport of Ions in Matter (TRIM) software for E$_{\rm IMP}$ = 25 -- 80~keV at a fixed fluence of $\Phi_{\rm Ga}$ = 4 $\times$ 10$^{\rm 16}$~cm$^{\rm -2}$. By lowering the implantation energy, $Ga$ distribution peak becomes narrower and taller as it shifts toward the top surface. Accordingly, three sets of Si:Ga samples with E$_{\rm IMP}$ = 25, 35, and 45~keV were prepared. At each implantation energy, samples underwent activation annealing at 350 to 800 $^{\circ}$C. Fig.\ref{fig.3}b displays the measured T$_{\rm c}$ a function of T$_{\rm DA}$ and E$_{\rm IMP}$, where lower energies on average lead to higher T$_{\rm c}$s, closer to values for $a$-Ga and $\gamma$-Ga \cite{campanini_raising_2018}. We note that superconductivity was absent in samples with E$_{\rm IMP}$ = 25~keV, and for E$_{\rm IMP}$ = 35~keV only annealing at 550~$^\circ$C yielded superconductivity. This is consistent with the TRIM simulation results for E$_{\rm IMP} <$ 45~keV; a larger portion of the $Ga$ peak resides within the SiO$_{\rm 2}$ cap (see Fig.\ref{fig.3}(a)), therefore the amounts that remain in Si post-annealed are not sufficient to form a connected $Ga$ network. Fig.\ref{fig.3}(c) shows the reentrant resistance R$_{\rm r}$ for the superconducting samples vs T$_{\rm DA}$ and E$_{\rm IMP}$. Increasing the annealing temperatures above 650~$^{\circ}$C, once again resulted in larger R$_{\rm r}$. Nonetheless, as seen in the inset of Fig.\ref{fig.3}(c), at T$_{\rm DA} <$ 650~$^{\circ}$C,  R$_{\rm r}$ was reduced by at least 5x when E$_{\rm IMP}$ smaller than 80~keV was used. More importantly, the sample with E$_{\rm IMP}$ = 45~keV \& T$_{\rm DA}$ = 575~$^{\circ}$C exhibited a near-zero R$_{\rm r}$, which can be ascribed to improved coupling between the superconducting $Ga$ clusters.

%%### FIGURE 3 ###############################
%%#########################################

\begin{figure}[hb]
\includegraphics[width=0.48\textwidth]{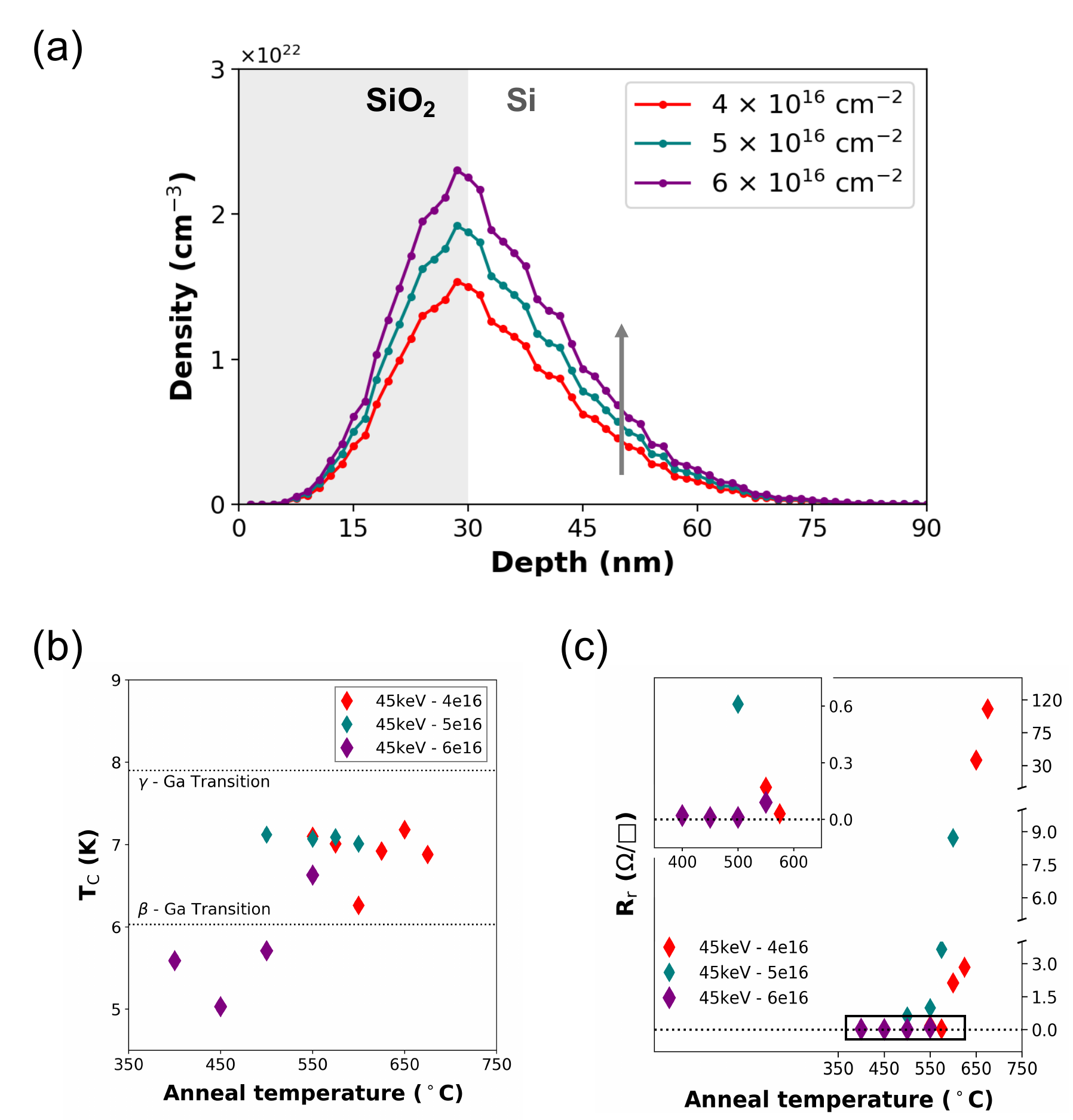}
\caption{\label{fig.4} (a) TRIM simulations of $Ga$ concentration vs depth at fixed implantation energy E$_{\rm IMP}$ of 45~keV when Ga$^+$ fluence $\Phi_{\rm Ga}$ varies between ($4 \times 10^{16}$~ cm$^{\rm -2}$ and $6 \times 10^{16}$~ cm$^{\rm -2}$). In all simulations, a 30~$nm$ thick SiO$_{\rm 2}$ is accounted for. (b) Transition temperatures T$_{\rm c}$ vs anneal temperature T$_{\rm {DA}}$ for all the superconducting Si:Ga samples with $\Phi_{\rm Ga}$ = ($4 \times 10^{16}$~ cm$^{\rm -2}$, $5 \times 10^{16}$~ cm$^{\rm -2}$, and $6 \times 10^{16}$~ cm$^{\rm -2}$). (c) Reentrant resistances at 1.5~K R$_{\rm r}$  vs T$_{\rm DA}$ and $\Phi_{\rm Ga}$. The inset magnifies the region outlined by the box near R$_{\rm r} \sim$ 0 ${\rm \Omega} / \square$.}
\end{figure}

Increasing the fluence during the implantation could further improve the connectivity of the $Ga$ clusters. In Fig.\ref{fig.4}(a), we show the TRIM simulations of implanted $Ga$ concentration vs depth for E$_{\rm IMP}$ = 45~keV and increasing $\Phi_{\rm Ga}$ from 4 x 10$^{16} cm^{\rm -2}$ to 6 $\times$ 10$^{\rm 16}$ cm$^{\rm -2}$. Increasing $\Phi_{\rm Ga}$ only raises the peak height while it leaves the peak position unchanged, thereby increasing the $Ga$ concentration per unit length within a narrow implanted Si region. Next, we prepared additional Si:Ga wafers with E$_{\rm IMP}$ = 45~keV and two fluence levels: $\Phi_{\rm Ga}$ = 5 x 10$^{16} cm^{\rm -2}$ and 6 $\times$ 10$^{\rm 16}$ cm$^{\rm -2}$. Fig.\ref{fig.4}(b) illustrates the T$_{\rm c}$ vs T$_{\rm DA}$ for superconducting Si:Ga specimens with $\Phi_{\rm Ga}$ = 4 -- 6 $\times$ 10$^{\rm 16}$ cm$^{\rm -2}$. Raising $\Phi_{\rm Ga}$ significantly shifts the T$_{\rm DA}$ window for superconductivity to lower temperatures. The shift is particularly noticeable for $\Phi_{\rm Ga}$ = 6 $\times$ 10$^{\rm 16}$ cm$^{\rm -2}$ where annealing at temperatures as low as 400~$^{\circ}$C has led to superconductivity. Additionally, majority of the T$_{\rm c}$ values at that fluence fell below $\beta-Ga$  T$_{\rm c}$, which may be a result of excess disorder. At high fluence, $Si$ and $O$ recoil events are more frequent throughout the implantation process leading to higher disorder in the superconducting phase. In Fig.\ref{fig.4}(c), we show the reentrant resistance R$_{\rm r}$ for the superconducting Si:Ga vs T$_{\rm DA}$ and $\Phi_{\rm Ga}$. We notice that increasing $\Phi_{\rm Ga}$ to 6 $\times$ 10$^{\rm 16}$ cm$^{\rm -2}$ significantly reduces the dependence of R$_{\rm r}$ on the T$_{\rm DA}$ while the absolute R$_{\rm r}$ values  at 1.5~K remain below 0.1 $\Omega / \square$.

 %%### FIGURE 4 ###############################
%%#########################################

\begin{figure}[hb!]
\includegraphics[width=0.45\textwidth]{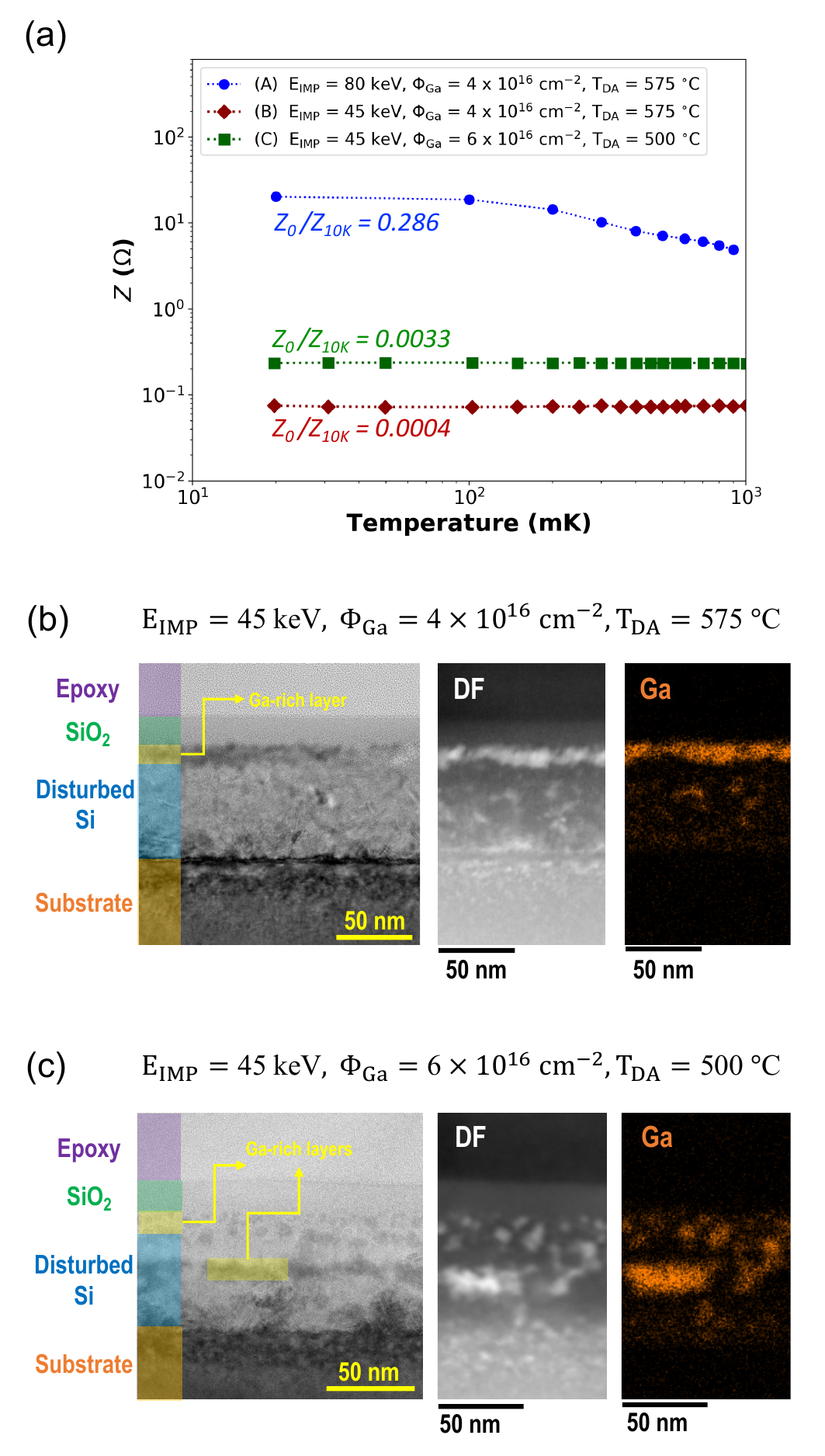}
\caption{\label{fig.5} (a) Temperature-dependent of AC impedance ($Z$) in 20~mK -- 1~K temperature range for three samples including: (A \textcolor{blue}{\CIRCLE}) E$_{\rm IMP}$ = 80~keV, $\Phi_{\rm Ga}$ = 4 $\times$ 10$^{\rm 16}$ cm$^{\rm -2}$, and T$_{\rm DA}$ = 575 $^{\circ}$C; (B \textcolor{BrickRed}{$\mdblkdiamond$}) E$_{\rm IMP}$ = 45~keV, $\Phi_{\rm Ga}$ = 4 $\times$ 10$^{\rm 16}$ cm$^{\rm -2}$, and T$_{\rm DA}$ = 575 $^{\circ}$C; (C \textcolor{ForestGreen}{$\mdblksquare$}) E$_{\rm IMP}$ = 45~keV, $\Phi_{\rm Ga}$ = 6 $\times$ 10$^{\rm 16}$ cm$^{\rm -2}$, and T$_{\rm DA}$ = 500 $^{\circ}$C. (b) TEM, STEM, and $Ga$ EDS maps for a Si samples with E$_{\rm IMP}$ = 45~keV, $\Phi_{\rm Ga}$ = 4 $\times$ 10$^{\rm 16}$ cm$^{\rm -2}$, and T$_{\rm DA}$ = 575 $^{\circ}$C. For ease of comparison zero-temperature impedance $Z_0$ normalized to  $Z_{10~K}$ is written for each sample. (b, c) Transmission electron microscopy, scanning dark-field (DF) and $Ga$ elemental maps for Si:Ga samples prepared in conditions indentical to samples (B) and (C), respectively.}
\end{figure}

Next, we turn to resistance measurements below 1~K carried out in a Triton dilution refrigerator (Oxford Instruments). Fig.\ref{fig.5}{a} displays the magnitude of AC impedance $Z$ vs temperature from 20~mK to 1~K for three samples with processing conditions that include: (A \textcolor{blue}{\CIRCLE}) E$_{\rm IMP}$ = 80~keV, $\Phi_{\rm Ga}$ = 4 $\times$ 10$^{\rm 16}$ cm$^{\rm -2}$, and T$_{\rm DA}$ = 575 $^{\circ}$C; (B \textcolor{BrickRed}{$\mdblkdiamond$}) E$_{\rm IMP}$ = 45~keV, $\Phi_{\rm Ga}$ = 4 $\times$ 10$^{\rm 16}$ cm$^{\rm -2}$, and T$_{\rm DA}$ = 575 $^{\circ}$C; (C \textcolor{ForestGreen}{$\mdblksquare$}) E$_{\rm IMP}$ = 45~keV, $\Phi_{\rm Ga}$ = 6 $\times$ 10$^{\rm 16}$ cm$^{\rm -2}$, and T$_{\rm DA}$ = 500 $^{\circ}$C. Samples (B) and (C) were chosen because they showed the lowest R$_{\rm r}$ at 1.5~K among all the samples presented in this study. In contrary, sample (A) represents a film with a clear reentrant behavior, although the smallest for samples with E$_{\rm IMP}$ = 80~keV. The reason for using $Z$ instead of R$_{\rm S}$ is the behavior of sample (A) below 1~K; the real part of $Z$ becomes negative while the imaginary part grows, consistent with an increasingly capacitive response. The $Z$ for (A) eventually reaches values as high as 20~$\Omega$ at 20~mK; about 29~\% of its normal impedance (Z$_{\rm 10K}$). In contrast, $Z$ for (B) and (C) are weakly temperature-dependent with negligible resistance down to 20~mK ( here $Z \approx R$).

%%### FIGURE 5 ###############################
%%#########################################

The TEM images and EDS maps displayed in Fig.\ref{fig.5}(b) \& (c) help connect the low-temperature behavior of samples (B) \& (C) to their structural characteristics, particularly their Ga cluster distributions. As a result of implantation at 45~keV, in both samples the depths of the disturbed regions have been reduced to 65 -- 70~$nm$. For sample (B) in shown Fig.\ref{fig.5}(b) Ga clusters appear to form a 8 -- 10~$nm$ thick densely-packed layer with little discontinuity along the Si/SiO$_{\rm 2}$ interface. For sample (C) shown in Fig.\ref{fig.5}, which has lower T$_{\rm DA}$ but higher fluence, $Ga$ clusters are instead dispersed within the top half of the disturbed zone alongside a Ga-rich bands that formed 25~$nm$ below the SiO$_{\rm 2}$ cap. Despite no clear evidence for continuity of $Ga$ cluster networks, absence of reentrant behavior in sample (C) confirms a stronger coherent coupling between the clusters. This could be partially explained by 1.5x higher dose of $Ga$ as dopant inside the Si weak-links effectively turning them into metals with no carrier freeze-out. Furthermore, the higher residual resistance for sample (C) relative to (B) may also be attributed to this difference in $Ga$ cluster coupling mechanism.

\setcitestyle{numbers}
%%%%################### Table 1 ####################################
%% Summary of B-T fit which  be changed after completing the WHH fit

\begin{table}[hb]
    \centering
    \caption{Summary of critical superconducting parameters for the three samples whose reentrant resistances were measured down to 20~mK. CC limit represents the Clogston--Chandrasekhar limit for the upper critical field defined in Ref.\cite{clogston_upper_1962}.}
    \begin{tabular}{|c|c|c|c|c|c|c|c|}
        \hline
        \# & E$_{\rm IMP}$ & $\Phi_{\rm Ga}$ & T$_{\rm DA}$  & $T_c$ & $B_{\rm 0}$ & CC Limit & $\xi_{\rm 0}$\\
        & & & & & & & \\
        & (keV) & (cm$^{\rm -2}$) & ($^{\circ}$C) & (K) & (T) & (T) & (nm)\\
        \hline
        (A) & 80 keV & 4 x 10$^{\rm 16}$ & 575 & 6.6 & 10.53 & 11.9 & 5.59\\
        \hline
        (B) & 45 keV & 4 x 10$^{\rm 16}$ & 575 & 7.0 & 12.06 & 12.6 & 5.22\\
        \hline
        (C) & 45 keV & 6 x 10$^{\rm 16}$ & 500 & 5.71 & 13.55 & 10.3 & 4.93\\
        \hline
    \end{tabular}
    \label{table.1}
\end{table}{}
%%%%################################################################# 
\setcitestyle{super}

Finally, to further evaluate the superconducting phases observed in Si:Ga samples, we studied their resistive transitions in magnetic fields. Table \ref{table.1} lists T$_{\rm c}$, critical magnetic field (B$_{\rm 0}$) and coherence length ($\xi_0$) for samples with processing conditions identical to (A), (B), and (C) in Fig.\ref{fig.5}(a). Additionally, the Clogston--Chandrasekhar (CC) limit \cite{clogston_upper_1962} is listed on the table for each sample. All samples exhibit large B$_{\rm 0}$'s consistent with filamentary type-II nature of superconductivity for which the CC limit was developed. In sample (C), however, the CC limit is surpassed by about 3.25~T. Moderate levels of deviation from CC limit in very thin lead films has been attributed to strong spin-orbit coupling in the 2D metal \cite{nam_ultrathin_2016}. Further studies will be underway to determine the possible roles of spin-orbit coupling and structural disorder in observation of such high critical fields.

%################################
% CONCLUSION
%################################

In conclusion, we have demonstrated the conditions to maintain zero resistance in hyperdoped Si:Ga down to millikelvin temperatures. We first found a reentrant resistive transition below T$_{\rm c}$ for samples prepared at higher implantation energies due to weak coupling between the superconducting $Ga$ clusters. By monitoring the reentrant resistance over a wide parameter space of E$_{\rm IMP}$ and $\Phi_Ga$, we targeted conditions that significantly improve the coherent coupling of $Ga$ cluster, therefore, eliminating the reentrant transition at temperatures as low as 20~mK. Our results should open a path for integration of hyperdoped Si:Ga into gate-tunable Josephson junctions as basic building blocks for functional superconducting circuits.\\\

%################################
% REFERENCE TO SUPPLEMENTAL DATA
%################################
We provide \textbf{supplementary material} that includes the optical images of Si:Ga surfaces, detailed resistance vs temperature traces, complementary TEM images and EDS line-scans of samples with 45~keV implantation energy, the magnetoresistance measurements, and the details of critical field and superconducting coherence length calculations.

%#####################################
% ACKNOWLEDGEMENT
%#####################################
\begin{acknowledgments}
The authors thank Matthieu Dartiailh for fruitful discussions. This research used resources of the Center for Functional Nanomaterials, which is a U.S. DOE Office of Science Facility, at Brookhaven National Laboratory under Contract No. DE-SC0012704. NYU team is supported by Intel, AFOSR Grant No. FA9550-16-1-0348, and ARO program “New and Emerging Qubit Science and Technology” Grant No. W911NF1810115. K.S. thanks Kroko, Inc for performing the ion implantation on the wafers. J.Y. acknowledges funding from the ARO QuaCGR fellowship reference No. W911NF1810067.
\end{acknowledgments}
%#####################################

% \nocite{*}
\bibliography{Si_SC_refs}% Produces the bibliography via BibTeX.

\begin{thebibliography}{21}%
\makeatletter
\providecommand \@ifxundefined [1]{%
 \@ifx{#1\undefined}
}%
\providecommand \@ifnum [1]{%
 \ifnum #1\expandafter \@firstoftwo
 \else \expandafter \@secondoftwo
 \fi
}%
\providecommand \@ifx [1]{%
 \ifx #1\expandafter \@firstoftwo
 \else \expandafter \@secondoftwo
 \fi
}%
\providecommand \natexlab [1]{#1}%
\providecommand \enquote  [1]{``#1''}%
\providecommand \bibnamefont  [1]{#1}%
\providecommand \bibfnamefont [1]{#1}%
\providecommand \citenamefont [1]{#1}%
\providecommand \href@noop [0]{\@secondoftwo}%
\providecommand \href [0]{\begingroup \@sanitize@url \@href}%
\providecommand \@href[1]{\@@startlink{#1}\@@href}%
\providecommand \@@href[1]{\endgroup#1\@@endlink}%
\providecommand \@sanitize@url [0]{\catcode `\\12\catcode `\$12\catcode
  `\&12\catcode `\#12\catcode `\^12\catcode `\_12\catcode `\%12\relax}%
\providecommand \@@startlink[1]{}%
\providecommand \@@endlink[0]{}%
\providecommand \url  [0]{\begingroup\@sanitize@url \@url }%
\providecommand \@url [1]{\endgroup\@href {#1}{\urlprefix }}%
\providecommand \urlprefix  [0]{URL }%
\providecommand \Eprint [0]{\href }%
\providecommand \doibase [0]{http://dx.doi.org/}%
\providecommand \selectlanguage [0]{\@gobble}%
\providecommand \bibinfo  [0]{\@secondoftwo}%
\providecommand \bibfield  [0]{\@secondoftwo}%
\providecommand \translation [1]{[#1]}%
\providecommand \BibitemOpen [0]{}%
\providecommand \bibitemStop [0]{}%
\providecommand \bibitemNoStop [0]{.\EOS\space}%
\providecommand \EOS [0]{\spacefactor3000\relax}%
\providecommand \BibitemShut  [1]{\csname bibitem#1\endcsname}%
\let\auto@bib@innerbib\@empty
%</preamble>
\bibitem [{\citenamefont {Shim}\ and\ \citenamefont
  {Tahan}(2014)}]{shim_bottom-up_2014}%
  \BibitemOpen
  \bibfield  {author} {\bibinfo {author} {\bibfnamefont {Y.-P.}\ \bibnamefont
  {Shim}}\ and\ \bibinfo {author} {\bibfnamefont {C.}~\bibnamefont {Tahan}},\
  }\bibfield  {title} {{\selectlanguage {en}\enquote {\bibinfo {title}
  {Bottom-up superconducting and {Josephson} junction devices inside a
  group-{IV} semiconductor},}\ }}\href {\doibase 10.1038/ncomms5225} {\bibfield
   {journal} {\bibinfo  {journal} {Nat Commun}\ }\textbf {\bibinfo {volume}
  {5}},\ \bibinfo {pages} {4225} (\bibinfo {year} {2014})}\BibitemShut
  {NoStop}%
\bibitem [{\citenamefont {Shim}\ and\ \citenamefont
  {Tahan}(2015)}]{shim_superconducting-semiconductor_2015}%
  \BibitemOpen
  \bibfield  {author} {\bibinfo {author} {\bibfnamefont {Y.-P.}\ \bibnamefont
  {Shim}}\ and\ \bibinfo {author} {\bibfnamefont {C.}~\bibnamefont {Tahan}},\
  }\bibfield  {title} {\enquote {\bibinfo {title}
  {Superconducting-{Semiconductor} {Quantum} {Devices}: {From} {Qubits} to
  {Particle} {Detectors}},}\ }\href {\doibase 10.1109/JSTQE.2014.2358208}
  {\bibfield  {journal} {\bibinfo  {journal} {IEEE Journal of Selected Topics
  in Quantum Electronics}\ }\textbf {\bibinfo {volume} {21}},\ \bibinfo {pages}
  {1--9} (\bibinfo {year} {2015})}\BibitemShut {NoStop}%
\bibitem [{\citenamefont {Bustarret}\ \emph {et~al.}(2006)\citenamefont
  {Bustarret}, \citenamefont {Marcenat}, \citenamefont {Achatz}, \citenamefont
  {Ka\v{c}mar\v{c}ik}, \citenamefont {L\'{e}vy}, \citenamefont {Huxley},
  \citenamefont {Ort\'{e}ga}, \citenamefont {Bourgeois}, \citenamefont {Blase},
  \citenamefont {D\'{e}barre},\ and\ \citenamefont
  {Boulmer}}]{bustarret_superconductivity_2006}%
  \BibitemOpen
  \bibfield  {author} {\bibinfo {author} {\bibfnamefont {E.}~\bibnamefont
  {Bustarret}}, \bibinfo {author} {\bibfnamefont {C.}~\bibnamefont {Marcenat}},
  \bibinfo {author} {\bibfnamefont {P.}~\bibnamefont {Achatz}}, \bibinfo
  {author} {\bibfnamefont {J.}~\bibnamefont {Ka\v{c}mar\v{c}ik}}, \bibinfo
  {author} {\bibfnamefont {F.}~\bibnamefont {L\'{e}vy}}, \bibinfo {author}
  {\bibfnamefont {A.}~\bibnamefont {Huxley}}, \bibinfo {author} {\bibfnamefont
  {L.}~\bibnamefont {Ort\'{e}ga}}, \bibinfo {author} {\bibfnamefont
  {E.}~\bibnamefont {Bourgeois}}, \bibinfo {author} {\bibfnamefont
  {X.}~\bibnamefont {Blase}}, \bibinfo {author} {\bibfnamefont
  {D.}~\bibnamefont {D\'{e}barre}}, \ and\ \bibinfo {author} {\bibfnamefont
  {J.}~\bibnamefont {Boulmer}},\ }\bibfield  {title} {{\selectlanguage
  {en}\enquote {\bibinfo {title} {Superconductivity in doped cubic silicon},}\
  }}\href {\doibase 10.1038/nature05340} {\bibfield  {journal} {\bibinfo
  {journal} {Nature}\ }\textbf {\bibinfo {volume} {444}},\ \bibinfo {pages}
  {465--468} (\bibinfo {year} {2006})}\BibitemShut {NoStop}%
\bibitem [{\citenamefont {Marcenat}\ \emph {et~al.}(2010)\citenamefont
  {Marcenat}, \citenamefont {Ka\v{c}mar\v{c}ík}, \citenamefont {Piquerel},
  \citenamefont {Achatz}, \citenamefont {Prudon}, \citenamefont {Dubois},
  \citenamefont {Gautier}, \citenamefont {Dupuy}, \citenamefont {Bustarret},
  \citenamefont {Ortega}, \citenamefont {Klein}, \citenamefont {Boulmer},
  \citenamefont {Kociniewski},\ and\ \citenamefont
  {D\'{e}barre}}]{marcenat_low-temperature_2010}%
  \BibitemOpen
  \bibfield  {author} {\bibinfo {author} {\bibfnamefont {C.}~\bibnamefont
  {Marcenat}}, \bibinfo {author} {\bibfnamefont {J.}~\bibnamefont
  {Ka\v{c}mar\v{c}ík}}, \bibinfo {author} {\bibfnamefont {R.}~\bibnamefont
  {Piquerel}}, \bibinfo {author} {\bibfnamefont {P.}~\bibnamefont {Achatz}},
  \bibinfo {author} {\bibfnamefont {G.}~\bibnamefont {Prudon}}, \bibinfo
  {author} {\bibfnamefont {C.}~\bibnamefont {Dubois}}, \bibinfo {author}
  {\bibfnamefont {B.}~\bibnamefont {Gautier}}, \bibinfo {author} {\bibfnamefont
  {J.~C.}\ \bibnamefont {Dupuy}}, \bibinfo {author} {\bibfnamefont
  {E.}~\bibnamefont {Bustarret}}, \bibinfo {author} {\bibfnamefont
  {L.}~\bibnamefont {Ortega}}, \bibinfo {author} {\bibfnamefont
  {T.}~\bibnamefont {Klein}}, \bibinfo {author} {\bibfnamefont
  {J.}~\bibnamefont {Boulmer}}, \bibinfo {author} {\bibfnamefont
  {T.}~\bibnamefont {Kociniewski}}, \ and\ \bibinfo {author} {\bibfnamefont
  {D.}~\bibnamefont {D\'{e}barre}},\ }\bibfield  {title} {{\selectlanguage
  {en}\enquote {\bibinfo {title} {Low-temperature transition to a
  superconducting phase in boron-doped silicon films grown on (001)-oriented
  silicon wafers},}\ }}\href {\doibase 10.1103/PhysRevB.81.020501} {\bibfield
  {journal} {\bibinfo  {journal} {Phys. Rev. B}\ }\textbf {\bibinfo {volume}
  {81}},\ \bibinfo {pages} {020501} (\bibinfo {year} {2010})}\BibitemShut
  {NoStop}%
\bibitem [{\citenamefont {Grockowiak}\ \emph {et~al.}(2013)\citenamefont
  {Grockowiak}, \citenamefont {Klein}, \citenamefont {Cercellier},
  \citenamefont {L\'{e}vy-Bertrand}, \citenamefont {Blase}, \citenamefont
  {Ka\v{c}mar\v{c}ik}, \citenamefont {Kociniewski}, \citenamefont {Chiodi},
  \citenamefont {D\'{e}barre}, \citenamefont {Prudon}, \citenamefont {Dubois},\
  and\ \citenamefont {Marcenat}}]{grockowiak_thickness_2013}%
  \BibitemOpen
  \bibfield  {author} {\bibinfo {author} {\bibfnamefont {A.}~\bibnamefont
  {Grockowiak}}, \bibinfo {author} {\bibfnamefont {T.}~\bibnamefont {Klein}},
  \bibinfo {author} {\bibfnamefont {H.}~\bibnamefont {Cercellier}}, \bibinfo
  {author} {\bibfnamefont {F.}~\bibnamefont {L\'{e}vy-Bertrand}}, \bibinfo
  {author} {\bibfnamefont {X.}~\bibnamefont {Blase}}, \bibinfo {author}
  {\bibfnamefont {J.}~\bibnamefont {Ka\v{c}mar\v{c}ik}}, \bibinfo {author}
  {\bibfnamefont {T.}~\bibnamefont {Kociniewski}}, \bibinfo {author}
  {\bibfnamefont {F.}~\bibnamefont {Chiodi}}, \bibinfo {author} {\bibfnamefont
  {D.}~\bibnamefont {D\'{e}barre}}, \bibinfo {author} {\bibfnamefont
  {G.}~\bibnamefont {Prudon}}, \bibinfo {author} {\bibfnamefont
  {C.}~\bibnamefont {Dubois}}, \ and\ \bibinfo {author} {\bibfnamefont
  {C.}~\bibnamefont {Marcenat}},\ }\bibfield  {title} {{\selectlanguage
  {en}\enquote {\bibinfo {title} {Thickness dependence of the superconducting
  critical temperature in heavily doped {Si}:{B} epilayers},}\ }}\href
  {\doibase 10.1103/PhysRevB.88.064508} {\bibfield  {journal} {\bibinfo
  {journal} {Phys. Rev. B}\ }\textbf {\bibinfo {volume} {88}},\ \bibinfo
  {pages} {064508} (\bibinfo {year} {2013})}\BibitemShut {NoStop}%
\bibitem [{\citenamefont {Skrotzki}\ \emph {et~al.}(2010)\citenamefont
  {Skrotzki}, \citenamefont {Fiedler}, \citenamefont {Herrmannsd\"{o}rfer},
  \citenamefont {Heera}, \citenamefont {Voelskow}, \citenamefont
  {M\"{u}cklich}, \citenamefont {Schmidt}, \citenamefont {Skorupa},
  \citenamefont {Gobsch}, \citenamefont {Helm},\ and\ \citenamefont
  {Wosnitza}}]{skrotzki_-chip_2010}%
  \BibitemOpen
  \bibfield  {author} {\bibinfo {author} {\bibfnamefont {R.}~\bibnamefont
  {Skrotzki}}, \bibinfo {author} {\bibfnamefont {J.}~\bibnamefont {Fiedler}},
  \bibinfo {author} {\bibfnamefont {T.}~\bibnamefont {Herrmannsd\"{o}rfer}},
  \bibinfo {author} {\bibfnamefont {V.}~\bibnamefont {Heera}}, \bibinfo
  {author} {\bibfnamefont {M.}~\bibnamefont {Voelskow}}, \bibinfo {author}
  {\bibfnamefont {A.}~\bibnamefont {M\"{u}cklich}}, \bibinfo {author}
  {\bibfnamefont {B.}~\bibnamefont {Schmidt}}, \bibinfo {author} {\bibfnamefont
  {W.}~\bibnamefont {Skorupa}}, \bibinfo {author} {\bibfnamefont
  {G.}~\bibnamefont {Gobsch}}, \bibinfo {author} {\bibfnamefont
  {M.}~\bibnamefont {Helm}}, \ and\ \bibinfo {author} {\bibfnamefont
  {J.}~\bibnamefont {Wosnitza}},\ }\bibfield  {title} {{\selectlanguage
  {en}\enquote {\bibinfo {title} {On-chip superconductivity via gallium
  overdoping of silicon},}\ }}\href {\doibase 10.1063/1.3509411} {\bibfield
  {journal} {\bibinfo  {journal} {Appl. Phys. Lett.}\ }\textbf {\bibinfo
  {volume} {97}},\ \bibinfo {pages} {192505} (\bibinfo {year}
  {2010})}\BibitemShut {NoStop}%
\bibitem [{\citenamefont {Fiedler}\ \emph {et~al.}(2011)\citenamefont
  {Fiedler}, \citenamefont {Heera}, \citenamefont {Skrotzki}, \citenamefont
  {Herrmannsd\"{o}rfer}, \citenamefont {Voelskow}, \citenamefont
  {M\"{u}cklich}, \citenamefont {Oswald}, \citenamefont {Schmidt},
  \citenamefont {Skorupa}, \citenamefont {Gobsch}, \citenamefont {Wosnitza},\
  and\ \citenamefont {Helm}}]{fiedler_superconducting_2011}%
  \BibitemOpen
  \bibfield  {author} {\bibinfo {author} {\bibfnamefont {J.}~\bibnamefont
  {Fiedler}}, \bibinfo {author} {\bibfnamefont {V.}~\bibnamefont {Heera}},
  \bibinfo {author} {\bibfnamefont {R.}~\bibnamefont {Skrotzki}}, \bibinfo
  {author} {\bibfnamefont {T.}~\bibnamefont {Herrmannsd\"{o}rfer}}, \bibinfo
  {author} {\bibfnamefont {M.}~\bibnamefont {Voelskow}}, \bibinfo {author}
  {\bibfnamefont {A.}~\bibnamefont {M\"{u}cklich}}, \bibinfo {author}
  {\bibfnamefont {S.}~\bibnamefont {Oswald}}, \bibinfo {author} {\bibfnamefont
  {B.}~\bibnamefont {Schmidt}}, \bibinfo {author} {\bibfnamefont
  {W.}~\bibnamefont {Skorupa}}, \bibinfo {author} {\bibfnamefont
  {G.}~\bibnamefont {Gobsch}}, \bibinfo {author} {\bibfnamefont
  {J.}~\bibnamefont {Wosnitza}}, \ and\ \bibinfo {author} {\bibfnamefont
  {M.}~\bibnamefont {Helm}},\ }\bibfield  {title} {{\selectlanguage
  {en}\enquote {\bibinfo {title} {Superconducting films fabricated by
  high-fluence {Ga} implantation in {Si}},}\ }}\href {\doibase
  10.1103/PhysRevB.83.214504} {\bibfield  {journal} {\bibinfo  {journal} {Phys.
  Rev. B}\ }\textbf {\bibinfo {volume} {83}},\ \bibinfo {pages} {214504}
  (\bibinfo {year} {2011})}\BibitemShut {NoStop}%
\bibitem [{\citenamefont {Heera}\ \emph {et~al.}(2013)\citenamefont {Heera},
  \citenamefont {Fiedler}, \citenamefont {H\"{u}bner}, \citenamefont {Schmidt},
  \citenamefont {Voelskow}, \citenamefont {Skorupa}, \citenamefont {Skrotzki},
  \citenamefont {Herrmannsd\"{o}rfer}, \citenamefont {Wosnitza},\ and\
  \citenamefont {Helm}}]{heera_silicon_2013}%
  \BibitemOpen
  \bibfield  {author} {\bibinfo {author} {\bibfnamefont {V.}~\bibnamefont
  {Heera}}, \bibinfo {author} {\bibfnamefont {J.}~\bibnamefont {Fiedler}},
  \bibinfo {author} {\bibfnamefont {R.}~\bibnamefont {H\"{u}bner}}, \bibinfo
  {author} {\bibfnamefont {B.}~\bibnamefont {Schmidt}}, \bibinfo {author}
  {\bibfnamefont {M.}~\bibnamefont {Voelskow}}, \bibinfo {author}
  {\bibfnamefont {W.}~\bibnamefont {Skorupa}}, \bibinfo {author} {\bibfnamefont
  {R.}~\bibnamefont {Skrotzki}}, \bibinfo {author} {\bibfnamefont
  {T.}~\bibnamefont {Herrmannsd\"{o}rfer}}, \bibinfo {author} {\bibfnamefont
  {J.}~\bibnamefont {Wosnitza}}, \ and\ \bibinfo {author} {\bibfnamefont
  {M.}~\bibnamefont {Helm}},\ }\bibfield  {title} {\enquote {\bibinfo {title}
  {Silicon films with gallium-rich nanoinclusions: from superconductor to
  insulator},}\ }\href {\doibase 10.1088/1367-2630/15/8/083022} {\bibfield
  {journal} {\bibinfo  {journal} {New J. Phys.}\ }\textbf {\bibinfo {volume}
  {15}},\ \bibinfo {pages} {083022} (\bibinfo {year} {2013})}\BibitemShut
  {NoStop}%
\bibitem [{\citenamefont {Chiodi}\ \emph {et~al.}(2017)\citenamefont {Chiodi},
  \citenamefont {Duvauchelle}, \citenamefont {Marcenat}, \citenamefont
  {D\'{e}barre},\ and\ \citenamefont
  {Lefloch}}]{chiodi_proximity-induced_2017}%
  \BibitemOpen
  \bibfield  {author} {\bibinfo {author} {\bibfnamefont {F.}~\bibnamefont
  {Chiodi}}, \bibinfo {author} {\bibfnamefont {J.-E.}\ \bibnamefont
  {Duvauchelle}}, \bibinfo {author} {\bibfnamefont {C.}~\bibnamefont
  {Marcenat}}, \bibinfo {author} {\bibfnamefont {D.}~\bibnamefont
  {D\'{e}barre}}, \ and\ \bibinfo {author} {\bibfnamefont {F.}~\bibnamefont
  {Lefloch}},\ }\bibfield  {title} {{\selectlanguage {en}\enquote {\bibinfo
  {title} {Proximity-induced superconductivity in all-silicon superconductor
  /normal-metal junctions},}\ }}\href {\doibase 10.1103/PhysRevB.96.024503}
  {\bibfield  {journal} {\bibinfo  {journal} {Phys. Rev. B}\ }\textbf {\bibinfo
  {volume} {96}},\ \bibinfo {pages} {024503} (\bibinfo {year}
  {2017})}\BibitemShut {NoStop}%
\bibitem [{\citenamefont {Duvauchelle}\ \emph {et~al.}(2015)\citenamefont
  {Duvauchelle}, \citenamefont {Francheteau}, \citenamefont {Marcenat},
  \citenamefont {Chiodi}, \citenamefont {D\'{e}barre}, \citenamefont
  {Hasselbach}, \citenamefont {Kirtley},\ and\ \citenamefont
  {Lefloch}}]{duvauchelle_silicon_2015}%
  \BibitemOpen
  \bibfield  {author} {\bibinfo {author} {\bibfnamefont {J.~E.}\ \bibnamefont
  {Duvauchelle}}, \bibinfo {author} {\bibfnamefont {A.}~\bibnamefont
  {Francheteau}}, \bibinfo {author} {\bibfnamefont {C.}~\bibnamefont
  {Marcenat}}, \bibinfo {author} {\bibfnamefont {F.}~\bibnamefont {Chiodi}},
  \bibinfo {author} {\bibfnamefont {D.}~\bibnamefont {D\'{e}barre}}, \bibinfo
  {author} {\bibfnamefont {K.}~\bibnamefont {Hasselbach}}, \bibinfo {author}
  {\bibfnamefont {J.~R.}\ \bibnamefont {Kirtley}}, \ and\ \bibinfo {author}
  {\bibfnamefont {F.}~\bibnamefont {Lefloch}},\ }\bibfield  {title}
  {{\selectlanguage {en}\enquote {\bibinfo {title} {Silicon superconducting
  quantum interference device},}\ }}\href {\doibase 10.1063/1.4928660}
  {\bibfield  {journal} {\bibinfo  {journal} {Appl. Phys. Lett.}\ }\textbf
  {\bibinfo {volume} {107}},\ \bibinfo {pages} {072601} (\bibinfo {year}
  {2015})}\BibitemShut {NoStop}%
\bibitem [{\citenamefont {Thorgrimsson}\ \emph {et~al.}(2020)\citenamefont
  {Thorgrimsson}, \citenamefont {McJunkin}, \citenamefont {MacQuarrie},
  \citenamefont {Coppersmith},\ and\ \citenamefont
  {Eriksson}}]{thorgrimsson_effect_2020}%
  \BibitemOpen
  \bibfield  {author} {\bibinfo {author} {\bibfnamefont {B.}~\bibnamefont
  {Thorgrimsson}}, \bibinfo {author} {\bibfnamefont {T.}~\bibnamefont
  {McJunkin}}, \bibinfo {author} {\bibfnamefont {E.~R.}\ \bibnamefont
  {MacQuarrie}}, \bibinfo {author} {\bibfnamefont {S.~N.}\ \bibnamefont
  {Coppersmith}}, \ and\ \bibinfo {author} {\bibfnamefont {M.~A.}\ \bibnamefont
  {Eriksson}},\ }\bibfield  {title} {\enquote {\bibinfo {title} {The effect of
  external electric fields on silicon with superconducting gallium
  nano-precipitates},}\ }\href {\doibase 10.1063/5.0002460} {\bibfield
  {journal} {\bibinfo  {journal} {Journal of Applied Physics}\ }\textbf
  {\bibinfo {volume} {127}},\ \bibinfo {pages} {215102} (\bibinfo {year}
  {2020})}\BibitemShut {NoStop}%
\bibitem [{\citenamefont {Fischer}\ \emph {et~al.}(2013)\citenamefont
  {Fischer}, \citenamefont {Pronin}, \citenamefont {Skrotzki}, \citenamefont
  {Herrmannsd\"{o}rfer}, \citenamefont {Wosnitza}, \citenamefont {Fiedler},
  \citenamefont {Heera}, \citenamefont {Helm},\ and\ \citenamefont
  {Schachinger}}]{fischer_optical_2013}%
  \BibitemOpen
  \bibfield  {author} {\bibinfo {author} {\bibfnamefont {T.}~\bibnamefont
  {Fischer}}, \bibinfo {author} {\bibfnamefont {A.~V.}\ \bibnamefont {Pronin}},
  \bibinfo {author} {\bibfnamefont {R.}~\bibnamefont {Skrotzki}}, \bibinfo
  {author} {\bibfnamefont {T.}~\bibnamefont {Herrmannsd\"{o}rfer}}, \bibinfo
  {author} {\bibfnamefont {J.}~\bibnamefont {Wosnitza}}, \bibinfo {author}
  {\bibfnamefont {J.}~\bibnamefont {Fiedler}}, \bibinfo {author} {\bibfnamefont
  {V.}~\bibnamefont {Heera}}, \bibinfo {author} {\bibfnamefont
  {M.}~\bibnamefont {Helm}}, \ and\ \bibinfo {author} {\bibfnamefont
  {E.}~\bibnamefont {Schachinger}},\ }\bibfield  {title} {{\selectlanguage
  {en}\enquote {\bibinfo {title} {Optical study of superconducting {Ga}-rich
  layers in silicon},}\ }}\href {\doibase 10.1103/PhysRevB.87.014502}
  {\bibfield  {journal} {\bibinfo  {journal} {Phys. Rev. B}\ }\textbf {\bibinfo
  {volume} {87}},\ \bibinfo {pages} {014502} (\bibinfo {year}
  {2013})}\BibitemShut {NoStop}%
\bibitem [{\citenamefont {Jaeger}\ \emph {et~al.}(1986)\citenamefont {Jaeger},
  \citenamefont {Haviland}, \citenamefont {Goldman},\ and\ \citenamefont
  {Orr}}]{jaeger_threshold_1986}%
  \BibitemOpen
  \bibfield  {author} {\bibinfo {author} {\bibfnamefont {H.~M.}\ \bibnamefont
  {Jaeger}}, \bibinfo {author} {\bibfnamefont {D.~B.}\ \bibnamefont
  {Haviland}}, \bibinfo {author} {\bibfnamefont {A.~M.}\ \bibnamefont
  {Goldman}}, \ and\ \bibinfo {author} {\bibfnamefont {B.~G.}\ \bibnamefont
  {Orr}},\ }\bibfield  {title} {\enquote {\bibinfo {title} {Threshold for
  superconductivity in ultrathin amorphous gallium films},}\ }\href {\doibase
  10.1103/PhysRevB.34.4920} {\bibfield  {journal} {\bibinfo  {journal} {Phys.
  Rev. B}\ }\textbf {\bibinfo {volume} {34}},\ \bibinfo {pages} {4920--4923}
  (\bibinfo {year} {1986})}\BibitemShut {NoStop}%
\bibitem [{\citenamefont {Hagel}\ \emph {et~al.}(2002)\citenamefont {Hagel},
  \citenamefont {Kelemen}, \citenamefont {Fischer}, \citenamefont {Pilawa},
  \citenamefont {Wosnitza}, \citenamefont {Dormann}, \citenamefont
  {L\"{o}hneysen}, \citenamefont {Schnepf}, \citenamefont {Schn\"{o}ckel},
  \citenamefont {Neisel},\ and\ \citenamefont
  {Beck}}]{hagel_superconductivity_2002}%
  \BibitemOpen
  \bibfield  {author} {\bibinfo {author} {\bibfnamefont {J.}~\bibnamefont
  {Hagel}}, \bibinfo {author} {\bibfnamefont {M.~T.}\ \bibnamefont {Kelemen}},
  \bibinfo {author} {\bibfnamefont {G.}~\bibnamefont {Fischer}}, \bibinfo
  {author} {\bibfnamefont {B.}~\bibnamefont {Pilawa}}, \bibinfo {author}
  {\bibfnamefont {J.}~\bibnamefont {Wosnitza}}, \bibinfo {author}
  {\bibfnamefont {E.}~\bibnamefont {Dormann}}, \bibinfo {author} {\bibfnamefont
  {H.~v.}\ \bibnamefont {L\"{o}hneysen}}, \bibinfo {author} {\bibfnamefont
  {A.}~\bibnamefont {Schnepf}}, \bibinfo {author} {\bibfnamefont
  {H.}~\bibnamefont {Schn\"{o}ckel}}, \bibinfo {author} {\bibfnamefont
  {U.}~\bibnamefont {Neisel}}, \ and\ \bibinfo {author} {\bibfnamefont
  {J.}~\bibnamefont {Beck}},\ }\bibfield  {title} {{\selectlanguage
  {en}\enquote {\bibinfo {title} {Superconductivity of a {Crystalline}
  {Ga84}-{Cluster} {Compound}},}\ }}\href {\doibase 10.1023/A:1020892022453}
  {\bibfield  {journal} {\bibinfo  {journal} {Journal of Low Temperature
  Physics}\ }\textbf {\bibinfo {volume} {129}},\ \bibinfo {pages} {133--142}
  (\bibinfo {year} {2002})}\BibitemShut {NoStop}%
\bibitem [{\citenamefont {Campanini}, \citenamefont {Diao},\ and\ \citenamefont
  {Rydh}(2018)}]{campanini_raising_2018}%
  \BibitemOpen
  \bibfield  {author} {\bibinfo {author} {\bibfnamefont {D.}~\bibnamefont
  {Campanini}}, \bibinfo {author} {\bibfnamefont {Z.}~\bibnamefont {Diao}}, \
  and\ \bibinfo {author} {\bibfnamefont {A.}~\bibnamefont {Rydh}},\ }\bibfield
  {title} {\enquote {\bibinfo {title} {Raising the superconducting
  $\textrm{T}_{c}$ of gallium: {In} situ characterization of the transformation
  of $\alpha$-{Ga} into $\beta$-{Ga}},}\ }\href {\doibase
  10.1103/PhysRevB.97.184517} {\bibfield  {journal} {\bibinfo  {journal} {Phys.
  Rev. B}\ }\textbf {\bibinfo {volume} {97}},\ \bibinfo {pages} {184517}
  (\bibinfo {year} {2018})}\BibitemShut {NoStop}%
\bibitem [{\citenamefont {Olesinski}, \citenamefont {Kanani},\ and\
  \citenamefont {Abbaschian}(1985)}]{olesinski_gasi_1985}%
  \BibitemOpen
  \bibfield  {author} {\bibinfo {author} {\bibfnamefont {R.~W.}\ \bibnamefont
  {Olesinski}}, \bibinfo {author} {\bibfnamefont {N.}~\bibnamefont {Kanani}}, \
  and\ \bibinfo {author} {\bibfnamefont {G.~J.}\ \bibnamefont {Abbaschian}},\
  }\bibfield  {title} {\enquote {\bibinfo {title} {The {Ga}-{Si}
  ({Gallium}-{Silicon}) system},}\ }\href@noop {} {\bibfield  {journal}
  {\bibinfo  {journal} {Bulletin of Alloy Phase Diagrams}\ }\textbf {\bibinfo
  {volume} {6}},\ \bibinfo {pages} {362--364} (\bibinfo {year}
  {1985})}\BibitemShut {NoStop}%
\bibitem [{\citenamefont {Nishino}, \citenamefont {Yamada},\ and\ \citenamefont
  {Kawabe}(1986)}]{nishino_carrier-concentration_1986}%
  \BibitemOpen
  \bibfield  {author} {\bibinfo {author} {\bibfnamefont {T.}~\bibnamefont
  {Nishino}}, \bibinfo {author} {\bibfnamefont {E.}~\bibnamefont {Yamada}}, \
  and\ \bibinfo {author} {\bibfnamefont {U.}~\bibnamefont {Kawabe}},\
  }\bibfield  {title} {\enquote {\bibinfo {title} {Carrier-concentration
  dependence of critical superconducting current induced by the proximity
  effect in silicon},}\ }\href {\doibase 10.1103/PhysRevB.33.2042} {\bibfield
  {journal} {\bibinfo  {journal} {Phys. Rev. B}\ }\textbf {\bibinfo {volume}
  {33}},\ \bibinfo {pages} {2042--2045} (\bibinfo {year} {1986})}\BibitemShut
  {NoStop}%
\bibitem [{\citenamefont {Kim}\ \emph {et~al.}(2001)\citenamefont {Kim},
  \citenamefont {Kim}, \citenamefont {Kang}, \citenamefont {Kim}, \citenamefont
  {Baranov}, \citenamefont {Park}, \citenamefont {Pshirkov},\ and\
  \citenamefont {Antipov}}]{kim_observation_2001}%
  \BibitemOpen
  \bibfield  {author} {\bibinfo {author} {\bibfnamefont {D.~C.}\ \bibnamefont
  {Kim}}, \bibinfo {author} {\bibfnamefont {J.~S.}\ \bibnamefont {Kim}},
  \bibinfo {author} {\bibfnamefont {H.~R.}\ \bibnamefont {Kang}}, \bibinfo
  {author} {\bibfnamefont {G.~T.}\ \bibnamefont {Kim}}, \bibinfo {author}
  {\bibfnamefont {A.~N.}\ \bibnamefont {Baranov}}, \bibinfo {author}
  {\bibfnamefont {Y.~W.}\ \bibnamefont {Park}}, \bibinfo {author}
  {\bibfnamefont {J.~S.}\ \bibnamefont {Pshirkov}}, \ and\ \bibinfo {author}
  {\bibfnamefont {E.~V.}\ \bibnamefont {Antipov}},\ }\bibfield  {title}
  {\enquote {\bibinfo {title} {Observation of anomalous reentrant
  superconductivity in
  \$\{{\textbackslash}mathrm\{{Sr}\}\}\_\{1{\textbackslash}ensuremath\{-\}x\}\{{\textbackslash}mathrm\{{K}\}\}\_\{x\}\{{\textbackslash}mathrm\{{BiO}\}\}\_\{3\}\$},}\
  }\href {\doibase 10.1103/PhysRevB.64.064502} {\bibfield  {journal} {\bibinfo
  {journal} {Phys. Rev. B}\ }\textbf {\bibinfo {volume} {64}},\ \bibinfo
  {pages} {064502} (\bibinfo {year} {2001})}\BibitemShut {NoStop}%
\bibitem [{\citenamefont {Ansermet}\ \emph {et~al.}(2016)\citenamefont
  {Ansermet}, \citenamefont {Petrović}, \citenamefont {He}, \citenamefont
  {Chernyshov}, \citenamefont {Hoesch}, \citenamefont {Salloum}, \citenamefont
  {Gougeon}, \citenamefont {Potel}, \citenamefont {Boeri}, \citenamefont
  {Andersen},\ and\ \citenamefont {Panagopoulos}}]{ansermet_reentrant_2016}%
  \BibitemOpen
  \bibfield  {author} {\bibinfo {author} {\bibfnamefont {D.}~\bibnamefont
  {Ansermet}}, \bibinfo {author} {\bibfnamefont {A.~P.}\ \bibnamefont
  {Petrović}}, \bibinfo {author} {\bibfnamefont {S.}~\bibnamefont {He}},
  \bibinfo {author} {\bibfnamefont {D.}~\bibnamefont {Chernyshov}}, \bibinfo
  {author} {\bibfnamefont {M.}~\bibnamefont {Hoesch}}, \bibinfo {author}
  {\bibfnamefont {D.}~\bibnamefont {Salloum}}, \bibinfo {author} {\bibfnamefont
  {P.}~\bibnamefont {Gougeon}}, \bibinfo {author} {\bibfnamefont
  {M.}~\bibnamefont {Potel}}, \bibinfo {author} {\bibfnamefont
  {L.}~\bibnamefont {Boeri}}, \bibinfo {author} {\bibfnamefont {O.~K.}\
  \bibnamefont {Andersen}}, \ and\ \bibinfo {author} {\bibfnamefont
  {C.}~\bibnamefont {Panagopoulos}},\ }\bibfield  {title} {\enquote {\bibinfo
  {title} {Reentrant {Phase} {Coherence} in {Superconducting} {Nanowire}
  {Composites}},}\ }\href {\doibase 10.1021/acsnano.5b05450} {\bibfield
  {journal} {\bibinfo  {journal} {ACS Nano}\ }\textbf {\bibinfo {volume}
  {10}},\ \bibinfo {pages} {515--523} (\bibinfo {year} {2016})},\ \bibinfo
  {note} {publisher: American Chemical Society}\BibitemShut {NoStop}%
\bibitem [{\citenamefont {Clogston}(1962)}]{clogston_upper_1962}%
  \BibitemOpen
  \bibfield  {author} {\bibinfo {author} {\bibfnamefont {A.~M.}\ \bibnamefont
  {Clogston}},\ }\bibfield  {title} {\enquote {\bibinfo {title} {Upper {Limit}
  for the {Critical} {Field} in {Hard} {Superconductors}},}\ }\href {\doibase
  10.1103/PhysRevLett.9.266} {\bibfield  {journal} {\bibinfo  {journal} {Phys.
  Rev. Lett.}\ }\textbf {\bibinfo {volume} {9}},\ \bibinfo {pages} {266--267}
  (\bibinfo {year} {1962})}\BibitemShut {NoStop}%
\bibitem [{\citenamefont {Nam}\ \emph {et~al.}(2016)\citenamefont {Nam},
  \citenamefont {Chen}, \citenamefont {Liu}, \citenamefont {Kim}, \citenamefont
  {Zhang}, \citenamefont {Yong}, \citenamefont {Lemberger}, \citenamefont
  {Kratz}, \citenamefont {Kirtley}, \citenamefont {Moler}, \citenamefont
  {Adams}, \citenamefont {MacDonald},\ and\ \citenamefont
  {Shih}}]{nam_ultrathin_2016}%
  \BibitemOpen
  \bibfield  {author} {\bibinfo {author} {\bibfnamefont {H.}~\bibnamefont
  {Nam}}, \bibinfo {author} {\bibfnamefont {H.}~\bibnamefont {Chen}}, \bibinfo
  {author} {\bibfnamefont {T.}~\bibnamefont {Liu}}, \bibinfo {author}
  {\bibfnamefont {J.}~\bibnamefont {Kim}}, \bibinfo {author} {\bibfnamefont
  {C.}~\bibnamefont {Zhang}}, \bibinfo {author} {\bibfnamefont
  {J.}~\bibnamefont {Yong}}, \bibinfo {author} {\bibfnamefont {T.~R.}\
  \bibnamefont {Lemberger}}, \bibinfo {author} {\bibfnamefont {P.~A.}\
  \bibnamefont {Kratz}}, \bibinfo {author} {\bibfnamefont {J.~R.}\ \bibnamefont
  {Kirtley}}, \bibinfo {author} {\bibfnamefont {K.}~\bibnamefont {Moler}},
  \bibinfo {author} {\bibfnamefont {P.~W.}\ \bibnamefont {Adams}}, \bibinfo
  {author} {\bibfnamefont {A.~H.}\ \bibnamefont {MacDonald}}, \ and\ \bibinfo
  {author} {\bibfnamefont {C.-K.}\ \bibnamefont {Shih}},\ }\bibfield  {title}
  {\enquote {\bibinfo {title} {Ultrathin two-dimensional superconductivity with
  strong spin–orbit coupling},}\ }\href {\doibase 10.1073/pnas.1611967113}
  {\bibfield  {journal} {\bibinfo  {journal} {Proc Natl Acad Sci U S A}\
  }\textbf {\bibinfo {volume} {113}},\ \bibinfo {pages} {10513--10517}
  (\bibinfo {year} {2016})}\BibitemShut {NoStop}%
\end{thebibliography}%

\end{document}